\documentclass[reqno]{amsart}

\usepackage{amsmath,amsfonts}
\usepackage{epsfig}
\usepackage{graphicx}

\def\alphaset{{\mathfrak A}}

\def\duh{{\rm Duh}}

\def\Gspace{{\mathfrak G}}

\def\opDelta{\widehat{\Delta}}
\def\opB{\widehat{B}}

\def\e{\epsilon}

\def\tr{{\rm Tr}}

\def\bra{\big\langle}
\def\ket{\big\rangle}

\def\N{{\mathbb N}}

\def\R{{\mathbb R}}

\def\ux{{\underline{x}}}

\def\cH{{\mathcal H}}

\def\1{{\bf 1}}

\def\eqnn{\begin{eqnarray*}}
\def\eeqnn{\end{eqnarray*}}
\def\eqn{\begin{eqnarray}}
\def\eeqn{\end{eqnarray}}

\def\prf{\begin{proof}}
\def\endprf{\end{proof}}

\theoremstyle{plain}
\newtheorem{theorem}{Theorem}[section]
\newtheorem{definition}[theorem]{Definition}

\newtheorem{lemma}[theorem]{Lemma}

\newtheorem{remark}[theorem]{Remark}

\numberwithin{equation}{section}

\begin{document}

\parskip=8pt

\title[Existence of solutions for the GP hierarchy]
{A new proof of existence of solutions for focusing and defocusing
Gross-Pitaevskii hierarchies}

\author[T. Chen]{Thomas Chen}
\address{T. Chen,  
Department of Mathematics, University of Texas at Austin.}
\email{tc@math.utexas.edu}

\author[N. Pavlovi\'{c}]{Nata\v{s}a Pavlovi\'{c}}
\address{N. Pavlovi\'{c},  
Department of Mathematics, University of Texas at Austin.}
\email{natasa@math.utexas.edu}


\begin{abstract}
We consider the cubic and quintic Gross-Pitaevskii (GP) hierarchies in $d\geq1$ dimensions, 
for focusing and defocusing interactions.  
We present a new proof of existence of solutions that does not require
the a priori bound on the spacetime norm, 
which was introduced in the work of Klainerman and Machedon, \cite{klma}, 
and used in our earlier work, \cite{chpa2}.
\end{abstract}

\maketitle

\section{Introduction}
\label{sec-modeldef-1}

In this note, we investigate the existence of solutions to the
Gross-Pitaevskii (GP) hierarchy, with focusing and defocusing interactions.  
We present a new proof of existence of solutions that does not require
the a priori bound on the spacetime norm used in our earlier work, \cite{chpa2},
which we adopted from the work of Klainerman and Machedon, \cite{klma}.

The GP hierarchy is a system of infinitely many coupled linear PDE's
describing a Bose gas of infinitely many
particles, interacting via delta interactions. Some GP hierarchies
(defocusing energy subcritical and focusing $L^2$-subcritical)
can be obtained as limits of BBGKY hierarchies of $N$-particle Schr\"odinger systems
of identical bosons, in the limit $N\rightarrow\infty$.
In the recent literature on this topic, 
there is a particular interest in the special class of factorized solutions to  GP hierarchies, 
which are parametrized
by solutions of a nonlinear Schr\"odinger (NLS) equation.
In this context, the NLS is interpreted as the mean field limit of an infinite system of interacting
bosons in the so-called Gross-Pitaevskii limit.
We refer to \cite{esy1,esy2,ey,kiscst,klma,rosc} and the references therein, 
and also to \cite{adgote,anasig,chpa,eesy,frgrsc,frknpi,frknsc,grma,grmama,he,sp}.
For recent mathematical developments focusing on the related problem of  Bose-Einstein 
condensation, we refer to
\cite{ailisesoyn,lise,lisesoyn,liseyn} and the references therein.

In a landmark series of works, Erd\"os, Schlein, and Yau \cite{esy1,esy2,ey}
provided the derivation of the cubic NLS as a dynamical mean field limit 
of an interacting Bose gas for a very general class of systems. 
The construction requires two main steps: 
\begin{enumerate} 
\item[(i)] Derivation of the GP hierarchy as the $N \rightarrow \infty$ limit
of the BBGKY hierarchy of density matrices associated to an $N$-body
Schr\"{o}dinger equation. 
The latter is defined for a scaling where the particle interaction potential tends to a delta distribution,
and where total kinetic and total interaction energy have the same order of magnitude in powers of $N$.
\item[(ii)] Proof of the uniqueness of solutions for the GP hierarchy.
It is subsequently verified that 
for factorized initial data, the solutions of the GP hierarchy are determined 
by a cubic NLS, for systems with 2-body interactions.
\end{enumerate} 
The proof of the uniqueness of solutions of the GP hierarchy 
is the most difficult part of this program, and it is obtained in \cite{esy1,esy2,ey}
by use of highly sophisticated Feynman graph expansion methods inspired by quantum 
field theory. 

In \cite{klma}, Klainerman and Machedon presented 
a different method to prove the uniqueness 
of solutions for the cubic GP hierarchy in $d=3$,  in a different space of solutions
than in \cite{esy1,esy2}.
Their approach uses Strichartz-type spacetime bounds on marginal density matrices, and a 
sophisticated combinatorial result,  obtained via a certain ``boardgame argument''
(which is a reformulation of a method developed in \cite{esy1,esy2}).
The analysis of Klainerman and Machedon requires the assumption of an a priori 
spacetime bound which is not proven in \cite{klma}. 
In \cite{kiscst}, Kirkpatrick, Schlein, and Staffilani proved that this a priori spacetime
bound is satisfied, locally in time, for the cubic GP hierarchy in $d=2$, 
by exploiting the conservation of
energy in the BBGKY hierarchy, in the limit as $N\rightarrow\infty$. 
In \cite{chpa}, we proved that the analogous a priori spacetime bound holds for the quintic GP
hierarchy in $d=1,2$.    

In \cite{chpa2}, we prove the existence and uniqueness of solutions in the 
spaces used by Klainerman and Machedon in \cite{klma}, and provide an estimate
that gives a precise meaning to their a priori assumption. 
For the proof, we introduce a natural topology on the space of sequences
of $k$-particle marginal density matrices
$$
	\Gspace \, = \, \{ \, \Gamma \, = \, ( \, \gamma^{(k)}(x_1,\dots,x_k;x_1',\dots,x_k') \, )_{k\in\N} 
	\, | \,
	\tr \gamma^{(k)} \, < \, \infty \, \}
$$
and invoke a contraction mapping argument.  
Accordingly, we prove in \cite{chpa2} local well-posedness for the cubic 
and quintic GP hierarchies, in various dimensions. 

In  \cite{chpa2}, we use the Klainerman-Machedon a priori assumption on the boundedness
of a certain spacetime norm for both the uniqueness and the existence parts of the proof
(which are obtained in the same step, via the contraction mapping argument).

In this paper, we give a new proof of the existence of solutions for 
focusing and defocusing $p$-GP hierarchies,  
without assuming any priori spacetime bounds.  
However, we prove as an a posteriori result that the Klainerman-Machedon spacetime bound is
indeed satisfied by this solution.

\subsection*{Organization of the paper} In Section \ref{sec-not} we introduce the 
GP hierarchy and the spaces that we use to analyze the hierarchy. In Section \ref{sec-main} we
state the main result of this paper, Theorem \ref{thm-localwp-TTstar-1-2},
and we present a proof of this theorem. The proof uses a free Strichartz estimate, as well as an 
iterated version of the Strichartz estimate, both of which are presented in Section \ref{sec-str}.

\section{The model} \label{sec-not}

In this section, we introduce the mathematical model analyzed in 
this paper. We will mostly adopt the notations and definitions from \cite{chpa2},
and we refer to \cite{chpa2} for motivations and more details.

\subsection{The spaces}
We introduce the space 
\eqn \nonumber 
	\Gspace \, := \, \bigoplus_{k=1}^\infty L^2(\R^{dk}\times\R^{dk})  
\eeqn
of sequences of density matrices
\eqn \nonumber 
	\Gamma \, := \, (\, \gamma^{(k)} \, )_{k\in\N}
\eeqn
where $\gamma^{(k)}\geq0$, $\tr\gamma^{(k)} =1$,
and where every $\gamma^{(k)}(\ux_k,\ux_k')$ is symmetric in all components of $\ux_k$,
and in all components of $\ux_k'$, respectively, i.e. 
\begin{equation}\label{symmetry}
	\gamma^{(k)}(x_{\pi (1)}, ...,x_{\pi (k)};x_{\pi'( 1)}^{\prime},
	 ...,x_{\pi'(k)}^{\prime})=\gamma^{(k)}( 	x_1, ...,x_{k};x_{1}^{\prime}, ...,x_{k}^{\prime})
\end{equation}
\\
holds for all $\pi,\pi'\in S_k$. 

For brevity, we will denote the vector $(x_1, \cdots, x_k)$ by 
$\ux_k$ and similarly the vector $(x'_1, \cdots, x'_k)$ by $\ux'_k$.
 
The $k$-particle marginals are assumed to be hermitean,
\begin{equation}\label{conj}
	\gamma^{(k)}(\ux_k;\ux_k')=\overline{\gamma^{(k)}(\ux_k';\ux_k) }.
\end{equation}
We call $\Gamma=(\gamma^{(k)})_{k\in\N}$ admissible if  
$\gamma^{(k)}=\tr_{k+1} \gamma^{(k+1)}$, that is,
\eqn 
	\gamma^{(k)}(\ux_k;\ux_k') 
	\, = \, \int dx_{k+1} \, \gamma^{(k+1)}(\ux_{k},x_{k+1};\ux_k',x_{k+1}) 
	\nonumber
\eeqn  
for all $k\in\N$.

Let $0<\xi<1$. We define
\eqn\label{eq-cHalpha-def-1} 
 	\cH_\xi^\alpha \, := \, \Big\{ \, \Gamma \, \in \, \Gspace \, \Big| \, \|\Gamma\|_{\cH_\xi^\alpha} < \, \infty \, \Big\}
\eeqn
where
\eqn \nonumber 
	\|\Gamma\|_{\cH_\xi^\alpha} \, = \, \sum_{k=1}^\infty \xi^{ k} 
	\| \,  \gamma^{(k)} \, \|_{H^\alpha_k(\R^{dk}\times\R^{dk})} \,,
\eeqn
with
\eqn\label{eq-gamma-norm-def-1}
	\| \gamma^{(k)} \|_{H^\alpha_k} & := & \big( \, \tr( \, | S^{(k,\alpha)}  \gamma^{(k)}|^2 \, ) \, \big)^{\frac12}
\eeqn 
where $S^{(k,\alpha)}:=\prod_{j=1}^k\bra\nabla_{x_j}\ket^\alpha\bra\nabla_{x_j'}\ket^\alpha$.

\subsection{The GP hierarchy} 
 
We introduce cubic, quintic, focusing, and defocusing GP hierarchies, 
using notations and definitions from \cite{chpa2}. 

Let $p\in\{2,4\}$.  The $p$-GP (Gross-Pitaevskii) hierarchy is given by
\eqn \label{eq-def-GP}
	i\partial_t \gamma^{(k)} \, = \, \sum_{j=1}^k [-\Delta_{x_j},\gamma^{(k)}]   
	\, + \,  \mu B_{k+\frac p2} \gamma^{(k+\frac p2)}
\eeqn
in $d$ dimensions, for $k\in\N$. Here,
\eqn \label{eq-def-b}
	B_{k+\frac p2}\gamma^{(k+\frac{p}{2})}
	\, = \, B^+_{k+\frac p2}\gamma^{(k+\frac{p}{2})}
        - B^-_{k+\frac p2}\gamma^{(k+\frac{p}{2})} \, ,
\eeqn
where 
$$ B^+_{k+\frac p2}\gamma^{(k+\frac{p}{2})}
   = \sum_{j=1}^k B^+_{j;k+1,\dots,k+\frac p2}\gamma^{(k+\frac{p}{2})},
$$ 
and 
$$ B^-_{k+\frac p2}\gamma^{(k+\frac{p}{2})}
   = \sum_{j=1}^k B^-_{j;k+1,\dots,k+\frac p2}\gamma^{(k+\frac{p}{2})},
$$                  
with 
\begin{align*} 
& \left(B^+_{j;k+1,\dots,k+\frac p2}\gamma^{(k+\frac{p}{2})}\right)
(t,x_1,\dots,x_k;x_1',\dots,x_k') \\
& \quad \quad = \int dx_{k+1}\cdots dx_{k+\frac p2} dx_{k+1}'\cdots dx_{k+\frac p2}' \\
& \quad\quad\quad\quad 
	\prod_{\ell=k+1}^{k+\frac p2} \delta(x_j-x_{\ell})\delta(x_j-x_{\ell}' )
        \gamma^{(k+\frac p2)}(t,x_1,\dots,x_{k+\frac p2};x_1',\dots,x_{k+\frac p2}'),
\end{align*} 
and 
\begin{align*} 
& \left(B^-_{j;k+1,\dots,k+\frac p2}\gamma^{(k+\frac{p}{2})}\right)
(t,x_1,\dots,x_k;x_1',\dots,x_k') \\
& \quad \quad = \int dx_{k+1}\cdots dx_{k+\frac p2} dx_{k+1}'\cdots dx_{k+\frac p2}' \\
& \quad\quad\quad\quad 
	\prod_{\ell=k+1}^{k+\frac p2} \delta(x'_j-x_{\ell})\delta(x'_j-x_{\ell}' )
        \gamma^{(k+\frac p2)}(t,x_1,\dots,x_{k+\frac p2};x_1',\dots,x_{k+\frac p2}').
\end{align*} 
The operator $B_{k+\frac p2}\gamma^{(k+\frac{p}{2})}$
accounts for  $\frac p2+1$-body interactions between the Bose particles.
We remark that for factorized solutions
\eqn\label{eq-GP-factorized-1}
\gamma^{(k)}(t,x_1,\dots,x_k;x_1',\dots,x_k') = 
\prod_{j=1}^{k} \phi(t, x_j) \, \bar{\phi}(t, x'_j),
\eeqn
the corresponding 1-particle wave function satisfies the
$p$-NLS 
$$i\partial_t\phi=-\Delta\phi+\mu|\phi|^p\phi$$ 
which is focusing if $\mu=-1$, and defocusing if $\mu=+1$.

As in \cite{chpa2, chpatz1}, we refer to (\ref{eq-def-GP}) 
as the {\em cubic GP hierarchy} if $p=2$,
and as the {\em quintic GP hierarchy} if $p=4$.
For $\mu=1$ or $\mu=-1$ we refer to the corresponding GP hierarchies as being  
defocusing or focusing, respectively.

The $p$-GP hierarchy can be rewritten in the following compact manner:
\eqn \label{chpa2-pGP}
        i\partial_t \Gamma \, + \, \opDelta_\pm \Gamma & = & \mu \opB \Gamma 
        \nonumber\\
        \Gamma(0) &=& \Gamma_0 \,,
\eeqn 
where
$$
	\opDelta_\pm \Gamma \, := \, ( \, \Delta^{(k)}_\pm \gamma^{(k)} \, )_{k\in\N} \, ,
        \; \; \; \; \mbox{ with }
        \Delta_{\pm}^{(k)} \, = \, \sum_{j=1}^{k} \left( \Delta_{x_j} - \Delta_{x'_j} \right)\, ,
$$
and 
\eqn \label{chpa2-B} 
	\opB \Gamma \, := \, ( \, B_{k+\frac{p}{2}} \gamma^{(k+\frac{p}{2})} \, )_{k\in\N} \,.
\eeqn
Also in this paper we will use the notation 
\begin{align*} 
	& \opB^+ \Gamma := \, ( \, B^+_{k+\frac{p}{2}} \gamma^{(k+\frac{p}{2})} \, )_{k\in\N}, 
        \nonumber \\  
	& \opB^- \Gamma := \, ( \, B^-_{k+\frac{p}{2}} \gamma^{(k+\frac{p}{2})} \, )_{k\in\N} \,.
\end{align*}
We refer to \cite{chpa2} for more detailed explanations.

\section{Statement and proof or the main Theorem} \label{sec-main} 

In our earlier work \cite{chpa2}, we proved local existence and uniqueness of solutions 
to the $p$-GP hierarchy in the space of solutions 
\eqn
	{\mathcal W}_\xi^\alpha(I) \, := \, \Big\{ \Gamma \, \Big| \, 
	\Gamma\in L^\infty_{t\in I}\cH_\xi^\alpha \; \; , \; \;
	\opB^+\Gamma \, , \, \opB^-\Gamma\in L^2_{t\in I}\cH_\xi^\alpha\Big\}
\eeqn
where $I:=[0,T]$. The requirement on the spacetime norm of $\opB^\pm\Gamma$ corresponds to the
Klainerman-Machedon a priori condition used in \cite{klma}.
For our proof, we used a contraction mapping argument, based on which both
existence and uniqueness were obtained in the same process. 
However, the question remained whether the condition that 
$\opB\Gamma\in L^2_{t\in I}\cH_\xi^\alpha$ for some $\xi$ is necessary for 
both the existence and uniqueness of solutions.
In the present paper, we prove that for the existence part, this a priori assumption is not required.
However, as an a posteriori result, we show that the solution obtained in this paper
has the property that $\opB\Gamma\in L^2_{t\in I}\cH_\xi^\alpha$.

We remark that for regularity $\alpha>\frac n2$, the result of this paper follows in an 
easier way by employing the estimate  
\eqn
	\|\opB\Gamma\|_{\cH_\xi^\alpha}\leq C\|\Gamma\|_{\cH_\xi^\alpha} 
	\label{eqn-Str-replace} 
\eeqn
instead of Strichartz estimates, which do not need to be invoked. 
The bound \eqref{eqn-Str-replace} for quintic GP was proved in our 
earlier work \cite{chpa} (see Theorem 4.3), and was 
employed to give a short proof of uniqueness for the quintic GP hierarchy (see Theorem 6.1 in \cite{chpa}). 
A bound of the type \eqref{eqn-Str-replace}  for the cubic GP was proved in a recent paper of 
Chen and Liu \cite{z_chli}, 
and was used to prove local well-posedness for the GP hierarchy in
the case when $\alpha>\frac n2$.

\begin{theorem} \label{thm-localwp-TTstar-1-2}
Let $\alpha \in \alphaset(d,p)$ where
\eqn
	\alphaset(d,p)
	\, := \,  \left\{
	\begin{array}{cc}
	(\frac12,\infty) & {\rm if} \; d=1 \\ 
	(\frac d2-\frac{1}{2(p-1)}, \infty) & {\rm if} \; d\geq2 \; {\rm and} \; (d,p)\neq(3,2)\\
	\big[1,\infty) & {\rm if} \; (d,p)=(3,2) \,,
	\end{array}
	\right.
\eeqn  
Assume that $\Gamma_0\in\cH_{\xi'}^\alpha$. Then, there exists a solution
of the $p$-GP hierarchy $\Gamma\in L^\infty_{t\in I}\cH_\xi^\alpha$ satisfying
\eqn
	\Gamma(t) \, = \, U(t)\Gamma_0 \, + \, i \, \mu\int_0^t U(t-s) \opB\Gamma(s) \, ds \,,
\eeqn
for $0<\xi<\xi'$ sufficiently small (it is sufficient that $\xi<\eta^2\xi'$ where the
constant $\eta$ is specified in Lemma \ref{lm-BGamma-Cauchy-1} below). 
\\
\\
In particular, this solution has the property that $\opB\Gamma\in L^2_{t\in I}\cH_\xi^\alpha$.
\end{theorem}

\begin{remark} 
We note that the presence of two different energy scales $\xi,\xi'$ has the
following interpretation on the level of the NLS. Let $R_0:=(\xi')^{-1/2}$ and
$R_1:=\xi^{-1/2}$.
Then, the local well-posedness result in Theorem \ref{thm-localwp-TTstar-1-2},
applied to
factorized initial data $\Gamma_0=\Gamma_{\phi_0}$ and the associated solution
$\Gamma(t)=\Gamma_{\phi(t)}$
(of the form \eqref{eq-GP-factorized-1}), is equivalent to the
following statement:
For $\|\phi_0\|_{H^1(\R^n)}<R_0$, there exists a unique solution
$\|\phi\|_{L^\infty_{t\in I}H^1(\R^n)}<R_1$, with $R_1>R_0$, in the space
$$\{\phi \in L^\infty_{t\in I}H^1(\R^n) \, | \,
\||\phi|^p\phi\|_{L^2_tH^1}<\infty\}\,.$$
This version of local well-posedness, specified  for  balls
$B_{R_0}(0),B_{R_1}(0)\subset H^1(\R^n)$,
contains the less specific formulation of local well-posedness where only
finiteness is required,
$\|\phi_0\|_{H^1(\R^n)}<\infty$ and $\|\phi\|_{L^\infty_{t\in
I}H^1(\R^n)}<\infty$.
\end{remark} 

\prf
The $p$-GP hierarchy is given by
\eqn \label{eq-def-GP-N-1}
	i\partial_t \gamma^{(n)} \, = \, \sum_{j=1}^n [-\Delta_{x_j},\gamma^{(n)}]   
	\, + \,  \mu B_{n+\frac p2} \gamma^{(n+\frac p2)}
\eeqn
for all $n\in\N$.

Let $P_{\leq N}$ denote the projection operator 
\eqn
	P_{\leq N} \; : \;  \Gspace &\rightarrow&\Gspace
	\nonumber\\
	\; \Gamma= (\gamma^{(1)},\gamma^{(2)},\dots)
	&\mapsto&(\gamma^{(1)},\dots,\gamma^{(N)},0,0,\dots) \,,
\eeqn
and $P_{>N}=1-P_{\leq N}$.
We consider the solution $\Gamma_N(t)$ of the  $p$-GP hierarchy,
\eqn
	i\partial_t\Gamma_N \, = \, \opDelta_\pm\Gamma_N \, + \, \mu \opB\Gamma_N \,,
\eeqn 
for the truncated initial data
\eqn 
	\Gamma_N(0) \, = \, P_{\leq N}\Gamma_0 \, = \, (\gamma_0^{(1)},\dots,\gamma_0^{(N)},0,0,\dots)
\eeqn
for an arbitrary, large, fixed $N\in\N$.

We observe that \eqref{eq-def-GP-N-1} determines a closed, infinite sub-hierarchy, 
for initial data $\gamma_N^{(n)}(0) =  0 $,  for $n > N$, which has the trivial solution
\eqn 
	\gamma_N^{(n)}(t) \, = \, 0 
	\; \; \; \; ,  \; \; \; t\in I=[0,T] \; \; \; , \; \; \; n>N \,.
\eeqn
Without invoking
uniqueness, it is not possible to conclude that this is the unique solution 
of the sub-hierarchy for  $n>N$ with zero initial data. 

However, for the
construction of a solution of  \eqref{eq-def-GP-N-1} (without any statement on uniqueness), 
we are free to choose $\gamma_N^{(n)}(t) = 0 $ for $n>N$.
In particular, it then follows that for $n \geq  N-\frac p2+1$, 
\eqn \label{eq-def-GP-N-finalterm-1}
	i\partial_t \gamma_N^{(n)} (t;\ux_N;\ux_N')
	\, = \, \sum_{j=1}^k [-\Delta_{x_j},\gamma_N^{(n)}] (t;\ux_N;\ux_N')  \,
\eeqn
solves the free evolution equation, since $B_{n+\frac p2} \gamma^{(n+\frac p2)}=0$, and thus,
\eqn 
	\gamma_N^{(n)}(t) \, = \, U^{(n)}(t) \, \gamma_N^{(n)}(0) \,.
\eeqn  
On the other hand, for $n\leq N-\frac p 2$, $\gamma_N^{(n)}(t)$ satisfies the $p$-GP hierarchy
in the full form \eqref{eq-def-GP-N-1}.

In conclusion, the solution of   \eqref{eq-def-GP-N-1} constructed above is given by
\eqn\label{eq-Duhamel-GammaN-1}
	\Gamma_N(t) \, = \, U(t)\Gamma_N(0) \, + \, i \, \mu\int_0^t U(t-s) \, \opB\Gamma_N(s) \, ds
\eeqn
for initial data $\Gamma_N(0)=P_{\leq N}\Gamma_0$.  

We introduce three parameters $\xi,\xi'',\xi'$ satisfying
\eqn
	\xi \, < \, \eta \, \xi'' \, < \, \eta^2 \, \xi' \,,
\eeqn
where the constant $0<\eta<1$ is  specified in Lemma \ref{lm-BGamma-Cauchy-1}
below.
Then it follows from Lemma \ref{lm-BGamma-Cauchy-1}  
that the sequence $(\opB \Gamma_N)_{N\in \N}$
is Cauchy in $L^2_{t\in I}\cH_{\xi''}^\alpha$. 
That is, for any $\e>0$, there exists $N(\e)\in\N$ such that  
\eqn 
	\| \opB( \Gamma_{N_1} - \Gamma_{N_2} )\|_{L^2_{t\in I}\cH_{\xi''}^\alpha}
	& \leq & C(\xi',\xi'') \,  \|P_{>N_1}\Gamma^{(n)}(0)\|_{\cH_{\xi'}^\alpha} 
	\nonumber\\
	&<&\e
	\label{eq-Gammadiff-Cauchy-aux-1}
\eeqn
holds for all $N_1,N_2>N(\e)$.
This is because by assumption, $\|\Gamma(0)\|_{\cH_{\xi'}^\alpha}<\infty$, which is a power series
in $\xi'>0$ with non-negative coefficients.
Hence, 
\eqn
	 \|P_{>N_1}\Gamma^{(n)}(0)\|_{\cH_{\xi'}^\alpha} \, = \,  \sum_{k >N_1} (\xi')^k \| \gamma^{(k)}(0) \|_{H^{\alpha}} \, \rightarrow \, 0
\eeqn 
as $N_1\rightarrow\infty$,
so that \eqref{eq-Gammadiff-Cauchy-aux-1} follows.

Accordingly, there exists a strong limit 
\eqn\label{eq-Thetaconv-1}
	\Theta \, = \, \lim_{N\rightarrow\infty}\opB \Gamma_N 
	\; \; \; \in \, L^2_{t\in I}\cH_{\xi''}^\alpha \,.
\eeqn
We claim that 
\eqn 
	\Theta(t) \, = \,  \opB U(t)\Gamma(0) \, + \, i \, \mu\int_0^t \opB U(t-s) \Theta(s) ds \,. \label{eqn-for-BGamma}
\eeqn
In order to prove this claim,  we observe that
\eqn
	\lefteqn{
	\Big\| \Theta(t) - \opB U(t)\Gamma(0)
	-i \, \mu\int_0^t \opB U(t-s) \Theta(s) \Big\|_{L^2_{t\in I} \cH_\xi^\alpha} 
	}
	\nonumber\\
	&\leq&\Big\|\Theta(t)-\opB \Gamma_N(t)\Big\|_{L^2_{t\in I} \cH_\xi^\alpha} 
	\label{eq-L2Hxi-Thetaconv-1-1}\\
	&& + \, 
	\Big\|\opB U(t)(\Gamma(0)- \Gamma_N(0))\Big\|_{L^2_{t\in I} \cH_\xi^\alpha} 
	\label{eq-L2Hxi-Thetaconv-1-2}\\
	&&+ \, \Big\|  \int_0^t ds \, \opB U(t-s) (\Theta(s)-\opB\Gamma_N(s)) 
	\Big\|_{L^2_{t\in I} \cH_\xi^\alpha} 
	\label{eq-L2Hxi-Thetaconv-1-3}\\
	&&+ \, 
	\Big\| \opB\Gamma_N(t) - \opB U(t)\Gamma_N(0)-i\int_0^t \opB U(t-s) \opB\Gamma_N(s) ds
	\Big\|_{L^2_{t\in I} \cH_\xi^\alpha} 
	\label{eq-L2Hxi-Thetaconv-1-4}\,.
\eeqn
First, we notice  that  \eqref{eq-L2Hxi-Thetaconv-1-4} is identically zero
because $\Gamma_N$ is a solution of the $p$-GP hierarchy, \eqref{eq-def-GP-N-1}.
Since $\xi < \xi''$ we can estimate the term \eqref{eq-L2Hxi-Thetaconv-1-1} as follows
\begin{align} 
	\Big\|\Theta(t)-\opB \Gamma_N(t)\Big\|_{L^2_{t\in I} \cH_\xi^\alpha} 
	& \leq  \|\Theta(t)-\opB \Gamma_N(t)\Big\|_{L^2_{t\in I} \cH_{\xi''}^\alpha} \nonumber \\
	= o_N(1) \label{bd-eq-L2Hxi-Thetaconv-1-1},
\end{align} 
where the last line follows from  \eqref{eq-Thetaconv-1}.
For  \eqref{eq-L2Hxi-Thetaconv-1-2}, since $\xi < \eta \, \xi''$
we can use the free Strichartz estimate \eqref{eq-freeStrichartz-L2H-1} as follows: 
\begin{align} 
	\Big\|\opB U(t)\left(\Gamma(0)- \Gamma_N(0)\right)\Big\|_{L^2_{t\in I} \cH_\xi^\alpha} 
	& \leq \|\Gamma(0)- \Gamma_N(0)\|_{\cH_{\xi''}^\alpha} \nonumber \\ 
	& \leq  \|\Gamma(0)- \Gamma_N(0)\|_{\cH_{\xi'}^\alpha} \nonumber \\ 
	& = \|P_{>N} \Gamma(0)\|_{\cH_{\xi'}^\alpha}. \label{bd-eq-L2Hxi-Thetaconv-1-2}
\end{align} 
For the term  \eqref{eq-L2Hxi-Thetaconv-1-3}, we have 
\eqn
 	\lefteqn{
	\Big\|  \int_0^t ds \, \opB U(t-s) (\Theta(s)-\opB\Gamma_N(s)) \Big\|_{L^2_{t\in I} \cH_\xi^\alpha} 
	}
	\nonumber\\
	&\leq&
	\Big\|  \int_0^t ds \, \Big\|\opB U(t-s) (\Theta(s)-\opB\Gamma_N(s)) \Big\|_{\cH_\xi^\alpha} 
	\Big\|_{L^2_{t\in I} }
	\nonumber\\
	&\leq&
	\int_0^T ds \, \Big\| \opB U(t-s) (\Theta(s)-\opB\Gamma_N(s)) \Big\|_{L^2_{t\in I} \cH_\xi^\alpha}
	\nonumber\\
	&\leq&
	C(T,\xi,\xi'') \, \int_0^T ds \Big\|  \Theta(s)-\opB\Gamma_N(s)  \Big\|_{ \cH_{\xi''}^\alpha}
	\label{eqn-towards-eqntheta-afterStr}\\
	&\leq&
	C(T,\xi,\xi'') \, T^{1/2} \,  \Big\|  \Theta(s)-\opB\Gamma_N(s)  \Big\|_{L^2_{t\in I} \cH_{\xi''}^\alpha} 
	\label{eqn-towards-eqntheta-afterHolder}\\ 
	&=&o_N(1) \,.
	\label{bd-eq-L2Hxi-Thetaconv-1-3}
\eeqn  
Here, to obtain \eqref{eqn-towards-eqntheta-afterStr} we use the free Strichartz estimate  \eqref{eq-freeStrichartz-L2H-1} 
in a manner similar to 
the $T-T^*$ argument for the Schr\"{o}dinger equation.
To obtain \eqref{eqn-towards-eqntheta-afterHolder} we used H\"{o}lder estimate and 
 to get the last line \eqref{bd-eq-L2Hxi-Thetaconv-1-3}, we used \eqref{eq-Thetaconv-1}.
In conclusion,
\eqn 
	\lefteqn{
	\Big\| \Theta(t) - \opB U(t)\Gamma(0)
	-i \, \mu \int_0^t \opB U(t-s) \Theta(s) \Big\|_{L^2_{t\in I} \cH_\xi^\alpha} 
	}
	\nonumber\\
	&&\hspace{1cm}\leq \,  o_N(1) \, + \, C(T,\xi) \,
	\|P_{>N}\Gamma(0) \|_{\cH_{\xi'}^\alpha}  
	\; \; \; \rightarrow \; 0 \; \; \; (N\rightarrow\infty) \,.
\eeqn
Therefore, taking the limit $N\rightarrow\infty$, we find that $\Theta$ satisfies
\eqn 
	\Theta(t) \, = \,  \opB U(t)\Gamma(0) \, + \, i \, \mu \int_0^t \opB U(t-s) \Theta(s) ds
\eeqn
as claimed.


Moreover, we observe that for $t \in I=[0,T]$ we have: 
\eqn
	\lefteqn{
	\|\Gamma_{N_1}(t)-\Gamma_{N_2}(t)\|_{\cH_\xi^\alpha}
	}
	\nonumber\\
	&\leq& \|U(t)(\Gamma_{N_1}(0)-\Gamma_{N_2}(0))\|_{\cH_\xi^\alpha}
	\, + \, 
	\|\int_0^t ds \, U(t-s) \opB(\Gamma_{N_1}(s)-\Gamma_{N_2}(s))\|_{\cH_\xi^\alpha}
	\nonumber\\
	&\leq&
	 \| \Gamma_{N_1}(0)-\Gamma_{N_2}(0)\|_{\cH_\xi^\alpha}
	\, + \, 
	\int_0^T ds \|  \opB(\Gamma_{N_1}(s)-\Gamma_{N_2}(s))\|_{\cH_\xi^\alpha}
	\nonumber\\
         &\leq&
	 \| \Gamma_{N_1}(0)-\Gamma_{N_2}(0)\|_{\cH_{\xi'}^\alpha}
	\, + \, 
	\int_0^T ds \|  \opB(\Gamma_{N_1}(s)-\Gamma_{N_2}(s))\|_{\cH_{\xi''}^\alpha}
	\label{eq-Glimit-xis}\\
	&\leq&
	 \| \Gamma_{N_1}(0)-\Gamma_{N_2}(0)\|_{\cH_{\xi'}^\alpha}
	\, + \, 
	T^{\frac12}  \|  \opB(\Gamma_{N_1}(s)-\Gamma_{N_2}(s))\|_{L^2_{s\in I}\cH_{\xi''}^\alpha}
	\nonumber\\
	&\leq&
	C(T,\xi',\xi'') \,
	 \| P_{>N_1} \Gamma_{0}\|_{\cH_{\xi'}^\alpha} \,, \nonumber
\eeqn
using the relation $\xi < \eta \xi'' < \eta^2 \xi'$ to obtain \eqref{eq-Glimit-xis}, 
and \eqref{eq-Gammadiff-Cauchy-aux-1} to pass to the last line. Thus, similarly
as in \eqref{eq-Gammadiff-Cauchy-aux-1}, there exists for every $\e>0$ a number $N(\e)\in\N$
such that
\eqn 
	\|\Gamma_{N_1}(t)-\Gamma_{N_2}(t)\|_{\cH_\xi^\alpha} \, < \, \e \,,
\eeqn
for all $N_1,N_2>N(\e)$. 
This implies that  $(\Gamma_N)_{N\in\N}$ is a Cauchy sequence in $L^{\infty}_{t \in I}\cH_\xi^\alpha$,
thus we obtain the strong limit
\eqn\label{eq-Gammaconv-1}
	\Gamma \, = \, \lim_{N\rightarrow\infty}\Gamma_N \; \; \in \; L^\infty_{t\in I}\cH_\xi^\alpha \,,
\eeqn
given the initial data $\Gamma_0\in\cH_{\xi'}^\alpha$.

Next, we claim that $\Gamma$ satisfies
\eqn\label{eq-Gamma-mildsol-1}
	\Gamma(t) \, = \, U(t) \Gamma_0 \, + \, i \, \mu \int_0^t U(t-s) \Theta(s) \, ds \,.
\eeqn
Indeed, we have for $t \in I$ that
\eqn
	\lefteqn{
	\Big\|\Gamma(t) \, - \,  U(t) \Gamma_0 
	\, - \, i \, \mu \int_0^t U(t-s) \Theta(s) \, ds \Big\|_{\cH_\xi^\alpha}
	}
	\nonumber\\
	&\leq&
	\|\Gamma(t)-\Gamma_N(t)\|_{\cH_\xi^\alpha}
	\label{eq-L2Hxi-Gammaconv-1-1}\\
	&&+ \, 
	\|U(t)(\Gamma_0-\Gamma_N(0))\|_{\cH_\xi^\alpha}
	\label{eq-L2Hxi-Gammaconv-1-2}\\
	&&+ \,
	\Big\|\int_0^tU(t-s)( \Theta(s) - \opB \Gamma_N(s)) \, ds \Big\|_{\cH_\xi^\alpha}
	\label{eq-L2Hxi-Gammaconv-1-3}\\
	&&+\,
	\Big\|\Gamma_N(t) \, - \,  U(t) \Gamma_N(0)
	\, - \, i \, \mu \int_0^t U(t-s)  \opB \Gamma_N(s) \, ds \Big\|_{\cH_\xi^\alpha} \,.
	\label{eq-L2Hxi-Gammaconv-1-4} 
\eeqn
Here, we note that \eqref{eq-L2Hxi-Gammaconv-1-4} is identically zero because 
$\Gamma_N(t)$ is a solution of the $p$-GP hierarchy,  \eqref{eq-def-GP-N-1}.
Moreover, we have 
\eqn
	\eqref{eq-L2Hxi-Gammaconv-1-3}
	&\leq&
	\int_0^T ds \Big\|  \Theta(s) - \opB \Gamma_N(s)  \Big\|_{\cH_\xi^\alpha}
	\nonumber\\
	&\leq&
	T^{\frac12} \, \Big\|  \Theta - \opB \Gamma_N  \Big\|_{L^2_{t\in I}\cH_\xi^\alpha}
	\nonumber\\
	&\leq&
	T^{\frac12} \, \Big\|  \Theta - \opB \Gamma_N  \Big\|_{L^2_{t\in I}\cH_{\xi''}^\alpha} \,.
\eeqn
Therefore,
 \eqn
	\lefteqn{
	\Big\|\Gamma(t) \, - \,  U(t) \Gamma_0 
	\, - \, i \, \mu\int_0^t U(t-s) \Theta(s) \, ds \Big\|_{L^\infty_{t\in I}\cH_\xi^\alpha}
	}
	\nonumber\\
	&\leq&
	\|\Gamma-\Gamma_N \|_{L_{t\in I}^\infty\cH_\xi^\alpha} \, + \, 
	\| \Gamma_0-\Gamma_N(0) \|_{\cH_\xi^\alpha} \, + \,
	T^{\frac12} \, \big\|  \Theta - \opB \Gamma_N  \big\|_{L^2_{t\in I}\cH_{\xi''}^\alpha} 
	\,, \; \; \; \;
\eeqn
where the right hand side
tends to zero as $N\rightarrow\infty$, due to the convergence
\eqref{eq-Gammaconv-1} and 
\eqref{eq-Thetaconv-1}.
This implies \eqref{eq-Gamma-mildsol-1}.

Finally, we observe that 
\eqn 
	\opB\Gamma(t) \, = \, \opB U(t) \Gamma_0 \, + \, i \, \mu \int_0^t \opB U(t-s) \Theta(s) \, ds
\eeqn
while
\eqn 
	\Theta(t) \, = \, \opB U(t) \Gamma_0 \, + \, i \, \mu \int_0^t \opB U(t-s) \Theta(s) \, ds \,.
\eeqn
Comparing the right hand sides, we infer that $\opB\Gamma=\Theta\in L^2_{t\in I}\cH_\xi^\alpha$.
This concludes the proof.
\endprf

\section{Iterated Duhamel formula and boardgame argument} \label{sec-str} 

In this section, it is our main goal to prove Lemma \ref{lm-BGamma-Cauchy-1} below.
We first summarize some results established in \cite{chpa,chpa2,klma},
which are related to Strichartz estimates for the GP hierarchy.

We first reformulate the Strichartz estimate for the free evolution 
$U(t)=e^{it\opDelta_\pm}=(U^{(n)}(t))_{n\in\N}$ proven in \cite{chpa2,klma}.

\begin{lemma}\label{lm-global-free-Strichartz-1}
Let $\alpha \in \alphaset(d,p)$.
Assume that $\Gamma_0\in\cH_{\xi'}^\alpha$ for some $0 < \xi' <1$.  
Then, for any $0<\xi < \xi'$, there exists a constant $C(\xi,\xi')$ such that 
the Strichartz estimate for the free evolution 
\eqn\label{eq-freeStrichartz-L2H-1}
	\|\opB U(t)\Gamma_0\|_{L_{t\in \R}^2\cH_{\xi }^\alpha} \,
	\leq \, C(\xi, \xi') \,   \|\Gamma_0\|_{\cH_{\xi'}^\alpha} \, 
\eeqn  
holds.
\end{lemma}

\prf 
From Theorem 1.3 in \cite{klma} and Proposition A.1 in \cite{chpa2}, 
we have, for $\alpha\in\alphaset(d,p)$, that
\eqn
	\lefteqn{
	\|B^{(k+\frac p2)}U^{(k+\frac p2)}(t)\gamma^{(k+\frac p2)}\|_{L^2_{t\in\R} H^\alpha_k}
	}
	\nonumber\\
	& \leq &2 \,
	\sum_{j=1}^k \, \| \, B_{j;k+1,\dots,k+\frac p2}^+  
	\, U^{(k+\frac p2)}(t)\gamma^{(k+\frac p2)} \, \|_{L^2_{t\in\R} H^\alpha_k}
	\nonumber\\
	& \leq & C \, k \,  \| \, \gamma^{(k+\frac p2)} \, \|_{H^\alpha_{k+\frac p2}} \,. 
	\label{eq-freeStrichartz-L2H-1-old}
\eeqn  
Then for any  $0 < \xi < \xi'$, we have: 
\begin{align}
	\|\opB U(t)\Gamma_0\|_{ L^2_{t\in\R}\cH_{\xi}^{\alpha} }
	& \leq 
	\sum_{k\geq1} \xi^k \|B^{(k+\frac p2)}U^{(k+\frac p2)}(t)
	\gamma^{(k+\frac p2)}\|_{L^2_{t\in\R} H^\alpha_k}
	\nonumber\\
	& \leq 
	C \, \sum_{k\geq1} k \, \xi^k \, \| \gamma_0^{(k+\frac p2)}\|_{H^\alpha_{k+\frac{p}{2}} } 
	\label{eq-Strprf-usefree}\\
	& = C \, ({\xi'})^{-\frac{p}{2}} \, \sum_{k\geq1} k \, \left( \frac{\xi}{\xi'} \right)^k \, 
	({\xi'})^{(k+\frac{p}{2})} \, \| \gamma_0^{(k+\frac p2)}\|_{H^\alpha_{k+\frac p2}} 
	\nonumber \\
	& \leq C \, ({\xi'})^{-\frac{p}{2}} \,  \sup_{k \geq 1} k \left( \frac{\xi}{\xi'}\right)^k \, \sum_{k\geq1} \, 
	({\xi'})^{(k+\frac{p}{2})} \, \| \gamma_0^{(k+\frac p2)}\|_{H^\alpha_{k+\frac p2}} 
	\nonumber \\
	& \leq C(\xi, \xi') \, \|\Gamma_0\|_{\cH_{\xi'}^\alpha} \, ,
	\nonumber 
\end{align} 
where to obtain \eqref{eq-Strprf-usefree} we used \eqref{eq-freeStrichartz-L2H-1-old}.
\endprf

\begin{definition}
Let
$\widetilde\Gamma=(\widetilde\gamma^{(n)})_{n\in\N}$ denote 
a sequence of arbitrary Schwartz class functions 
$\widetilde\gamma^{(n)}\in{\mathcal S}(\R\times\R^{nd}\times\R^{nd})$.
Then, we define the associated sequence $\duh_j(\widetilde\Gamma)$ 
of  {\em $j$-th level iterated Duhamel terms}, with components given by
\eqn\label{eq-Duh-j-def-1}
	\lefteqn{
	\duh_j(\widetilde\Gamma)^{(n+\frac p2)}(t) 
	}
	\\
	& := & (-i\mu)^j\int_0^t dt_1 \cdots \int_0^{t_{j-2}}dt_{j-1}
	e^{i(t-t_1)\Delta_\pm^{(n+\frac p2)}}B_{n+\frac {2p}2}e^{i(t_1-t_2)\Delta_\pm^{(n+\frac {2p}2)}}
	\nonumber\\
	&&\quad\quad\quad\quad\quad\quad
	B_{n+\frac {3p}2} \cdots
	\cdots B_{n+\frac {jp}2} e^{i t_{j-1} \Delta_\pm^{(n+\frac {jp}2)}} 
	\widetilde\gamma^{(n+\frac {jp}2)}(t_{j-1}) \,. \;\;
	\nonumber
\eeqn 
\end{definition}

As usual, the definition is given for Schwartz class functions, and can be extended
to the spaces in discussion by density arguments. The fact that 
$\duh_j(\widetilde\Gamma)^{(n)}\in{\mathcal S}(\R\times\R^{nd}\times\R^{nd})$
holds under the above conditions, for all $n$,
can be easily verified.
Using the boardgame strategy of \cite{klma} (which is a reformulation of a 
combinatorial argument developed in \cite{esy1,esy2}), one obtains:

\begin{lemma}
\label{lm-boardgame-est-1}
Let $\alpha\in\alphaset(d,p)$ and $p\in\{2,4\}$. 
Then, for $\widetilde\Gamma=(\widetilde\gamma^{(n)})_{n\in\N}$ as above,
\eqn\label{eq-BGamma-Duhj-combin-bd-1}
	\lefteqn{
	\| \, B_{n+\frac p2}\duh_j(\widetilde\Gamma)^{(n+\frac p2)}(t) 
	\, \|_{L^2_{t\in I}H^\alpha(\R^{nd}\times\R^{nd})} 
	}
	\\
	&&\hspace{1cm}
	\, \leq \, n C_0^n (c_0 T)^{\frac {j}2} \|B_{n+\frac{jp}{2}}U^{(n+\frac p2)}(\,\cdot\,)
	\widetilde\gamma^{(n+\frac{jp}2)}
	\|_{L^2_{t\in I}H^\alpha(\R^{(n+\frac{jp}2)d}\times\R^{(n+\frac{jp}2)d})}  \,,
	\nonumber
\eeqn 
where the constants $c_0,C_0$ depend only on $d,p$.
\end{lemma}

For the proof in the cubic case, $p=2$, we refer to \cite{chpa2,klma}, and for the 
proof in the quintic case, $p=4$, to \cite{chpa}.

We then observe that any solution $\Gamma_N$ of \eqref{eq-def-GP-N-1} with initial data
$\Gamma_N(0)=P_{\leq N}\Gamma_0$ satisfies the equation
(obtained from acting with $\opB$ on \eqref{eq-Duhamel-GammaN-1})
\eqn
	\opB\Gamma_N(t) \, = \, \opB U(t) \Gamma_N(0) \, + \, i \int_0^t \opB U(t-s) \opB\Gamma_N(s) ds
\eeqn
and by iteration,
\eqn
	(\opB\Gamma_N)^{(n)}(t) & = & \sum_{j=1}^{k-1}B_{n+\frac p2}
	\duh_j(\Gamma_N(0))^{(n+\frac p2)}(t)
	\nonumber\\
	&&\hspace{2cm}
	\, + \, B_{n+\frac p2}\duh_{k}(\opB\Gamma_N)^{(n+\frac p2)}(t) \,,
\eeqn
obtained from iterating the Duhamel formula $k$ times for the $n$-th component of $\opB\Gamma$.
Since $\Gamma_N^{(m)}(t)=0$ for all $m>N$, the remainder term on the last line is zero
whenever $n+\frac{kp}{2}>N$. Thus,
 \eqn
	(\opB\Gamma_N)^{(n)}(t) & = & \sum_{j=1}^{\lceil 2(N-n)/p\rceil}B_{n+\frac p2}
	\duh_j(\Gamma_N(0))^{(n+\frac p2)}(t)  \,,
\eeqn
where each term on the right explicitly depends only on the initial data $\Gamma_N(0)$
(there is no implicitly dependence on the solution $\Gamma_N(t)$).

\begin{lemma}
\label{lm-BGamma-Cauchy-1}
Assume that $\alpha\in\alphaset(d,p)$ and $p\in\{2,4\}$, and that 
$\Gamma_0\in\cH_{\xi'}^\alpha$ for some $0<\xi'<1$.
Let $N_1,N_2\in\N$, where $N_1<N_2$. Then, there exists a constant $0<\eta=\eta(d,p)<1$ such that the estimate
\eqn 
	\| \opB( \Gamma_{N_1} - \Gamma_{N_2} )\|_{L^2_{t\in I}\cH_\xi^\alpha}
	\, \leq \, C(T,\xi,\xi') \,  \|P_{>N_1}\Gamma_0\|_{\cH_{\xi'}^\alpha} 
	\label{eq-Gammadiff-Cauchy-aux-2}
\eeqn
holds whenever $\xi<\eta\xi'$ (we note that it suffices to let $\eta<C_0^{-1}$ where the
constant $C_0=C_0(d,p)$ is specified in Lemma \ref{lm-boardgame-est-1}).
\end{lemma}

\prf
For simplicity of notation, we shall present the explicit arguments 
for the (cubic) case $p=2$.
The (quintic) case $p=4$ is completely analogous.

We have
 \eqn\label{eq-BGammaN1minN2-1-1}
	(\opB(\Gamma_{N_1}-\Gamma_{N_2}))^{(n)}(t) & = & \sum_{j=1}^{ N_1-n}B_{n+1}
	\duh_j((\Gamma_{N_1}(0)-\Gamma_{N_2}(0))^{(n+1)}(t) 
	\nonumber\\
	&&- \, \sum_{j=N_1-n+1}^{ N_2-n}B_{n+1}
	\duh_j(\Gamma_{N_2}(0))^{(n+1)}(t)  \,,
	\label{eq-BGammaN1minN2-1-2}
\eeqn
using the fact that $\gamma_{N_1}^{(n+j)}=0$ for $j>N_1-n$, see \eqref{eq-Duh-j-def-1}.

Since $\gamma_{N_1}^{(n)}(0)=\gamma_{N_2}^{(n)}(0)$ for $1\leq n\leq N_1$, the
first sum on the rhs of \eqref{eq-BGammaN1minN2-1-1} is identically zero.

For the second term on the rhs of \eqref{eq-BGammaN1minN2-1-2}, we have for 
the summation index that $j\geq N_1-n+1$. Thus, the components of $\Gamma_{N_2}(0)$
occurring in \eqref{eq-BGammaN1minN2-1-2} are given by $\gamma_{N_2}^{(n+j)}$ 
with $j\geq N_1-n+1$, that is,  $\gamma_{N_2}^{(m)}$ with $m>N_1$.

Using  Lemma \ref{lm-boardgame-est-1} and the free Strichartz estimate \eqref{eq-freeStrichartz-L2H-1-old},
we therefore find that
\eqn  
	\lefteqn{
	\|(\opB(\Gamma_{N_1}-\Gamma_{N_2}))^{(n)}(t) \|_{L^2_{t\in I}H^\alpha}
	}
	\nonumber\\
	& \leq &   \sum_{j=N_1-n+1}^{ N_2-n}
	\|B_{n+1}
	\duh_j(\Gamma_{N_2}(0))^{(n+1)}(t)\|_{L^2_{t\in I}H^\alpha}
	\nonumber\\
	& \leq &   \sum_{j=N_1-n+1}^{ N_2-n}
	n C_0^n (c_0T)^{\frac j2} \|B_{n+j}U^{(n+j)}(t)\gamma_{N_2}^{n+j}(0)\|_{L^2_{t\in I}H^\alpha}
	\nonumber\\
	& \leq & ( \xi')^{-n} 
	n^2 \, C_0^n
	\sum_{j=N_1-n+1}^{ N_2-n} 
	(c_0T(\xi')^{-2})^{\frac j2}(\xi')^{n+j} \| \gamma_{N_2}^{n+j}(0)\|_{ H^\alpha}
	\nonumber\\
	& \leq &  (\xi')^{-n} 
	n^2 \, C_0^n \, C_1(T,\xi') \, \| P_{>N_1}\Gamma_{N_2} (0)\|_{ \cH_{\xi'}^\alpha}
	 \,,
	\label{eq-BGammaN1minN2-2}
\eeqn
for $T>0$ sufficiently small so that $c_0T(\xi')^{-2}\leq1$.
Hence, 
\eqn  
	\lefteqn{
	\sum_{n\in\N}\xi^n\|(\opB(\Gamma_{N_1}-\Gamma_{N_2}))^{(n)}(t) \|_{L^2_{t\in I}H^\alpha}
	}
	\nonumber\\
	& \leq &     C_1(T,\xi') 
	\Big(\sum_{n\in\N} n^2 \, C_0^n \, (\xi/\xi')^n \Big)
	\, \| P_{>N_1}\Gamma_{N_2} (0)\|_{ \cH_{\xi'}^\alpha}
	\nonumber\\
	& \leq & C(T,\xi,\xi') \,    \| P_{>N_1}\Gamma_{N_2} (0)\|_{ \cH_{\xi'}^\alpha}
	 \,,
	\label{eq-BGammaN1minN2-3}
\eeqn
for $\xi<\eta\xi'$ where $\eta<C_0^{-1}$, noting that $C_0=C_0(d,p)$.

This proves the claim for the case $p=2$.
The case $p=4$ is completely analogous, and we shall omit a repetition of arguments.
\endprf
 
\subsection*{Acknowledgements}
We are grateful to Igor Rodnianski for pointing out an error in an earlier version of this work,
and for useful comments.
The work of T.C. is supported by NSF grant DMS 0704031 / DMS-0940145 and DMS-1009448.
The work of N.P. is supported by NSF grant number DMS 0758247 
and an Alfred P. Sloan Research Fellowship.

\end{document}